\documentclass[twocolumn]{aastex631}

\newcommand{\OIII}{[\textrm{O}~\textsc{iii}]\xspace}
\newcommand{\Ha}{\textrm{H}\ensuremath{\alpha}\xspace}

\newcommand{\cii}{{\rm\,[C{\sc II}]}}

\usepackage{xspace} 
\usepackage{amsmath}
\usepackage{CJK}
\usepackage{multirow}
\usepackage{gensymb}
\graphicspath{ {./figures/} }

\begin{document}
\begin{CJK*}{UTF8}{gbsn}

\title{MAGNIF: A Tentative Lensed Rotating Disk at $z=8.34$ detected by JWST NIRCam WFSS with Dynamical Forward Modeling
}

\author[0000-0001-5951-459X]{Zihao Li (黎子豪)}
\affiliation{Department of Astronomy, Tsinghua University, Beijing 100084, China; \href{mailto:li-zh21@mails.tsinghua.edu.cn}{\textnormal{li-zh21@mails.tsinghua.edu.cn}}}
\author[0000-0001-8467-6478]{Zheng Cai}
\affiliation{Department of Astronomy, Tsinghua University, Beijing 100084, China; \href{mailto:li-zh21@mails.tsinghua.edu.cn}{\textnormal{li-zh21@mails.tsinghua.edu.cn}}}
\author[0000-0002-4622-6617]{Fengwu Sun} 
\affiliation{Steward Observatory, University of Arizona, 933 N Cherry Avenue, Tucson, AZ 85721, USA}
\author[0000-0001-5492-1049 ]{Johan Richard}
\affiliation{Univ Lyon, Univ Lyon1, Ens de Lyon, CNRS, Centre de Recherche Astrophysique de Lyon UMR5574, F-69230, Saint-Genis-Laval, France}
\author[0000-0002-6849-5375]{Maxime Trebitsch}
\affiliation{Kapteyn Astronomical Institute, University of Groningen, 9700 AV Groningen, The Netherlands}
\author[0000-0003-4337-6211]{Jakob M. Helton}
\affiliation{Steward Observatory, University of Arizona, 933 N Cherry Avenue, Tucson, AZ 85721, USA}
\author[0000-0001-9065-3926]{Jose M. Diego}
\affiliation{Instituto de Física de Cantabria (CSIC-UC). Avenida Los Castros s/n. E-39005 Santander, Spain}
\author[0000-0003-3484-399X]{Masamune Oguri}
\affiliation{Center for Frontier Science, Chiba University, 1-33 Yayoi-cho, Inage-ku, Chiba 263-8522, Japan}
\affiliation{Department of Physics, Graduate School of Science, Chiba University, 1-33 Yayoi-Cho, Inage-Ku, Chiba 263-8522, Japan}
\author[0000-0002-7460-8460]{Nicholas Foo}
\affiliation{School of Earth and Space Exploration, Arizona State University,
Tempe, AZ 85287-1404, USA}
\author[0000-0001-6052-4234]{Xiaojing Lin}
\affiliation{Department of Astronomy, Tsinghua University, Beijing 100084, China; \href{mailto:li-zh21@mails.tsinghua.edu.cn}{\textnormal{li-zh21@mails.tsinghua.edu.cn}}}
\author[0000-0002-8686-8737]{Franz Bauer}
\affiliation{Instituto de Astrof{\'{\i}}sica and Centro de Astroingenier{\'{\i}}a, Facultad de F{\'{i}}sica, Pontificia Universidad Cat{\'{o}}lica de Chile, Campus San Joaquín, Av. Vicuña Mackenna 4860, Macul Santiago, 7820436, Chile}
\affiliation{Millennium Institute of Astrophysics, Nuncio Monse{\~{n}}or S{\'{o}}tero Sanz 100, Of 104, Providencia, Santiago, Chile}
\affiliation{Space Science Institute, 4750 Walnut Street, Suite 205, Boulder, Colorado 80301, USA}
\author[0000-0002-3805-0789]{Chian-Chou Chen}
\affiliation{Academia Sinica Institute of Astronomy and Astrophysics (ASIAA), No. 1, Sec. 4, Roosevelt Road, Taipei 10617, Taiwan}
\author[0000-0003-1949-7638]{Christopher J. Conselice}
\affiliation{Jodrell Bank Centre for Astrophysics, University of Manchester, Oxford Road, Manchester UK}
\author[0000-0002-8726-7685]{Daniel Espada}
\affiliation{Departamento de F\'{i}sica Te\'{o}rica y del Cosmos, Campus de Fuentenueva, Edificio Mecenas, Universidad de Granada, E-18071, Granada, Spain}
\affiliation{Instituto Carlos I de F\'{i}sica Te\'{o}rica y Computacional, Facultad de Ciencias, E-18071, Granada, Spain}
\author[0000-0003-1344-9475]{Eiichi Egami}
\affiliation{Steward Observatory, University of Arizona, 933 N Cherry Avenue, Tucson, AZ 85721, USA}
\author[0000-0003-3310-0131]{Xiaohui Fan}
\affiliation{Steward Observatory, University of Arizona, 933 N Cherry Avenue, Tucson, AZ 85721, USA}
\author[0000-0003-1625-8009]{Brenda L.~Frye}
\affiliation{Steward Observatory, University of Arizona, 933 N Cherry Avenue, Tucson, AZ 85721, USA}
\author[0000-0001-7440-8832]{Yoshinobu Fudamoto}
\affiliation{Waseda Research Institute for Science and Engineering, Faculty of Science and Engineering, Waseda University, 3-4-1 Okubo, Shinjuku, Tokyo 169-8555, Japan}
\affiliation{National Astronomical Observatory of Japan, 2-21-1, Osawa, Mitaka, Tokyo, Japan}
\author[0000-0003-4528-5639]{Pablo G. P\'erez-Gonz\'alez}
\affiliation{Centro de Astrobiolog\'{\i}a (CAB), CSIC-INTA, Ctra. de Ajalvir km 4, Torrej\'on de Ardoz, E-28850, Madrid, Spain}
\author[0000-0003-4565-8239]{Kevin Hainline}
\affiliation{Steward Observatory, University of Arizona, 933 N Cherry Avenue, Tucson, AZ 85721, USA}
\author[0000-0003-4512-8705]{Tiger Yu-Yang Hsiao}
\affiliation{Department of Physics and Astronomy, The Johns Hopkins University, 3400 N Charles St. Baltimore, MD 21218, USA}
\author[0000-0001-7673-2257]{Zhiyuan Ji}
\affiliation{Steward Observatory, University of Arizona, 933 N Cherry Avenue, Tucson, AZ 85721, USA}
\author[0000-0002-5768-738X]{Xiangyu Jin}
\affiliation{Steward Observatory, University of Arizona, 933 N Cherry Avenue, Tucson, AZ 85721, USA}
\author[0000-0002-6610-2048]{Anton M. Koekemoer}
\affiliation{Space Telescope Science Institute, 3700 San Martin Dr., Baltimore, MD 21218, USA}
\author[0000-0002-5588-9156]{Vasily Kokorev}
\affiliation{Kapteyn Astronomical Institute, University of Groningen, 9700 AV Groningen, The Netherlands}
\author[0000-0002-4052-2394]{Kotaro Kohno}
\affiliation{Institute of Astronomy, Graduate School of Science, The University of Tokyo, 2-21-1 Osawa, Mitaka, Tokyo 181-0015, Japan}
\affiliation{Research Center for the Early Universe, Graduate School of Science, The University of Tokyo, 7-3-1 Hongo, Bunkyo-ku, Tokyo 113-0033, Japan}
\author[0000-0001-6251-649X]{Mingyu Li}
\affiliation{Department of Astronomy, Tsinghua University, Beijing 100084, China; \href{mailto:li-zh21@mails.tsinghua.edu.cn}{\textnormal{li-zh21@mails.tsinghua.edu.cn}}}
\author[0000-0002-2419-3068]{Minju Lee}
\affiliation{Cosmic Dawn Center (DAWN), Jagtvej 128, DK2200 Copenhagen N, Denmark}
\affiliation{DTU-Space, Technical University of Denmark, Elektrovej 327, DK2800 Kgs. Lyngby, Denmark}
\author[0000-0002-4872-2294]{Georgios E. Magdis}
\affiliation{Cosmic Dawn Center (DAWN), Jagtvej 128, DK2200 Copenhagen N, Denmark}
\affiliation{DTU-Space, Technical University of Denmark, Elektrovej 327, DK2800 Kgs. Lyngby, Denmark}
\affiliation{Niels Bohr Institute, University of Copenhagen, Jagtvej 128, DK2200 Copenhagen N, Denmark}

\author[0000-0001-9262-9997]{Christopher N. A. Willmer}
\affiliation{Steward Observatory, University of Arizona, 933 N Cherry Avenue, Tucson, AZ 85721, USA}
\author[0000-0001-8156-6281]{Rogier A. Windhorst}
\affiliation{School of Earth and Space Exploration, Arizona State University,
Tempe, AZ 85287-1404, USA}
\author[0000-0003-0111-8249]{Yunjing Wu}
\affiliation{Department of Astronomy, Tsinghua University, Beijing 100084, China; \href{mailto:li-zh21@mails.tsinghua.edu.cn}{\textnormal{li-zh21@mails.tsinghua.edu.cn}}}
\affiliation{Steward Observatory, University of Arizona, 933 N Cherry Avenue, Tucson, AZ 85721, USA}
\author[0000-0001-7592-7714]{Haojing Yan}
\affiliation{Department of Physics and Astronomy, University of Missouri, Columbia, MO 65211, USA}
\author[0000-0002-4321-3538]{Haowen Zhang (张昊文)}
\affiliation{Steward Observatory, University of Arizona, 933 N Cherry Avenue, Tucson, AZ 85721, USA}
\author[0000-0002-0350-4488]{Adi Zitrin}
\affiliation{Physics Department, Ben-Gurion University of the Negev, P.O. Box 653, Be'er-Sheva 84105, Israel}
\author[0000-0002-3983-6484]{Siwei Zou}
\affiliation{Department of Astronomy, Tsinghua University, Beijing 100084, China; \href{mailto:li-zh21@mails.tsinghua.edu.cn}{\textnormal{li-zh21@mails.tsinghua.edu.cn}}}
\author[0000-0002-8630-6435]{Fuyan Bian}
\affiliation{European Southern Observatory, Alonso de C´ordova 3107, Casilla 19001, Vitacura, Santiago 19, Chile}
\author[0000-0003-0202-0534]{Cheng Cheng}
\affiliation{Chinese Academy of Sciences South America Center for Astronomy, National Astronomical Observatories, CAS, Beijing 100101, China}
\author[0000-0002-4781-9078]{Christa DeCoursey}
\affiliation{Steward Observatory, University of Arizona, 933 N Cherry Avenue, Tucson, AZ 85721, USA}
\author[0000-0001-6278-032X]{Lukas J. Furtak}
\affiliation{Physics Department, Ben-Gurion University of the Negev, P.O. Box 653, Be'er-Sheva 84105, Israel}
\author[0000-0003-3780-6801]{Charles Steinhardt}
\affiliation{Cosmic Dawn Center (DAWN), Jagtvej 128, DK2200 Copenhagen N, Denmark}
\affiliation{DTU-Space, Technical University of Denmark, Elektrovej 327, DK2800 Kgs. Lyngby, Denmark}
\affiliation{Niels Bohr Institute, University of Copenhagen, Jagtvej 128, DK2200 Copenhagen N, Denmark}
\author[0000-0003-1937-0573]{Hideki Umehata}
\affiliation{Institute for Advanced Research, Nagoya University, Furocho, Chikusa, Nagoya 464-8602, Japan}
\affiliation{Department of Physics, Graduate School of Science, Nagoya University, Furocho, Chikusa, Nagoya 464-8602, Japan}
\affiliation{Cahill Center for Astronomy and Astrophysics, California Institute of Technology, MS 249-17, Pasadena, CA 91125, USA}


\begin{abstract}
We report galaxy MACS0416-Y3 behind the lensing cluster MACSJ0416.1--2403 as a tentative rotating disk at $z=8.34$ detected through its $\OIII\lambda5007$ emission in JWST NIRCam wide-field slitless spectroscopic observations. The discovery is based on our new grism dynamical modeling methodology for JWST NIRCam slitless spectroscopy, using the data from ``Median-band Astrophysics with the Grism of NIRCam in Frontier Fields'' (MAGNIF), a JWST Cycle-2 program. The $\OIII\lambda5007$ emission line morphology in grism data shows velocity offsets compared to the F480M direct imaging, suggestive of rotation. Assuming a geometrically thin disk model, we constrain the rotation velocity of $v_{\rm rot}=58^{+53}_{-35}$ km s$^{-1}$ via forward modeling of the two-dimensional (2D) spectrum. We obtain the kinematic ratio of $v_{\rm rot}/\sigma_v=1.6^{+1.9}_{-0.9}$, where $\sigma_v$ is the velocity dispersion, in line with a quasi-stable thin disk. The resulting dynamical mass is estimated to be $\log(M_{\rm dyn}/M_{\odot})=8.4^{+0.5}_{-0.7}$. If the rotation confirmed, our discovery suggests that rotating gaseous disks may have already existed within 600 million years after Big Bang.

\end{abstract}

\keywords{Galaxy dynamics (591), Galaxy evolution (594), Galaxy chemical evolution (580), High redshift galaxies (734)}

\section{Introduction}\label{sec:intro}

An outstanding question in galaxy evolution is the time at which galactic rotating disks are formed. 
In the cosmological model of $\Lambda$-Cold Dark Matter ($\Lambda$CDM), a bottom-up structure build-up is favored. 
Gaseous material and 
dark matter feed into dark matter halos through accretion or merger, 
allowing the structure growth \citep{Lin_65,Zel'dovich_70}.
Nevertheless, detailed physical processes dominating galaxy 
formation are still hotly debated.
For massive galaxies, it is believed that the infalling gas is shock-heated to the virial temperature and accretes 
spherically before cooling and condensing into a disk 
that is sustained by rotation in the so called  ``hot mode" \citep{Dekel_06,Joung_12,Hafen_22}. In this scenario, the disk of gas and 
stars form relatively late.
In addition to ``hot mode" accretion, 
numerical simulations predict an additional scenario 
in which gas efficiently accretes onto galaxies with low halo masses through flows along filamentary structures, with a sizeable portion 
of the gas remaining cool, at temperatures much below 
the virial temperature of the galaxies \citep{Keres_05,Ocvirk_08,Dekel_09}. 
In contrast to the former models, within this scenario, 
disk galaxies could be established at an early stage, as streams of cold gas from the intergalactic medium (IGM) could directly form a rotating disk as the gas spirals inward \citep{Martin_16}, and such cold streams are responsible for the formation of disks at high redshift \citep{Dekel_09a}.
Numerical simulations show that star formation at $z>6$ is mostly fueled by the efficient accretion of cold gas \citep{Yajima_15}, and hot mode makes negligible contribution to star formation since most of the shock-heated gas cannot cool within a Hubble time \citep{van_11b}. To discriminate between these mass accretion models and learn how galaxies acquire their mass, one has to trace the earliest onset of galaxy disks (either via gas or stars, e.g., \citealt{Ubler_22}).

Carbon monoxide (CO) or UV/optical spectroscopy at $z\approx 2.5$ 
has been used to identify disks \citep[e.g.,][]{Aravena_14,Genzel_17}. 
High angular-momentum cold-accretion on the larger scale of 
the circumgalactic medium (CGM) has also been identified at $z\approx2$ \citep{Zhang_23}. 
Observations on CO with JVLA and \cii\ with ALMA at higher redshifts of $z=4-5$ 
have produced suggestive evidence of cold-gas disks being supported by rotation \citep{Neeleman_20,Oliveira_23}. 

It is still hard to detect definitive rotating disk galaxies at $z>5$, caused by a combination of relatively low resolution and 
sensitivity. 
Most of the successful detections from the ground were obtained by ALMA \citep{Smit_18,Rizzo_20,Fujimoto_21,Lelli_21,Pope_23,Posses_23}, but the rotation features are not always observed \citep{Tamura_23}. While it is relatively easier to detect the rotating disks among bright quasar host galaxies \citep{Pensabene_20,Izumi_21,Neeleman_21},  less massive systems are less commonly detected and often require the aid from gravitational lensing \citep{Fujimoto_21}. The most distant disk candidate so far resides at $z=9.1$ with $\OIII\ 88\mu m$ detected in a gravitationally lensed galaxy, MACS1149-JD1 \citep{Hashimoto_18,Tokuoka_22}. 

Many hydrodynamical simulations predict ordered rotation of cold gas at redshifts as high as $z\sim10$ \citep{Katz_19}. With the launch of JWST, there are emerging evidences that the disk galaxies are more frequent than expected by previous HST observations at redshift up to $z\sim8$ \citep{Ferreira_22,Ferreira_23}. Such high redshift disks are expected to be detected by JWST through optical emission lines (e.g. \Ha, \OIII), which trace warm/ionized gas, in complement to cold gas traced by CO or the more complex phases traced by \cii in radio bands. The kinematics of the multi-phase ISM may greatly help us to understand the early galaxy formation and mass assembly.

In this letter, we study the galaxy MACS0416-Y3 \citep{Coe_15,Infante_15,Laporte_15,McLeod_15} behind the lensing cluster MACSJ0416.1--2403 and report it as a possible rotating disk at $z>8$. 
The discovery is based on our methodology to forward model the kinematics of rotating disks. We find its kinematic ratio (ratio between rotation velocity and velocity dispersion) can be well predicted by semi-empirical models \citep{Wisnioski_15}, indicating a quasi-stable gas disk which has formed less than 600 Myr after the Big Bang.

The letter is organized as follows. We describe our observations and detail the data reduction procedure in Section \ref{sec:data}. We present the lens modeling and dynamical modeling in Section \ref{sec:analysis}. The kinematic results are presented in Section \ref{sec:res} and we conclude our findings in Section \ref{sec:con}.
Throughout this letter, we adopt the AB magnitude system \citep{Oke_83}, and assume a flat $\Lambda$CDM cosmology with $\Omega_m=0.3,\ \Omega_\Lambda=0.7$, and $H_0=70\ {\rm km\ s^{-1} Mpc}^{-1}$. We use the following vacuum line wavelengths: $4960.295$\AA\ for $\OIII\lambda4959$ and $5008.240$\AA\ for $\OIII\lambda5007$ based on the Atomic Line List v2.04\footnote{\url{https://linelist.pa.uky.edu/atomic/index.html}}.

\section{Observations and Data reduction}\label{sec:data}
The JWST NIRCam imaging and wide field slitless spectrograph (WFSS) data were obtained through the Cycle-2 General Observer (GO) program ``Median-band Astrophysics with the Grism of NIRCam in Frontier Fields'' (MAGNIF; PID: 2883, PI: F.\ Sun).
The detailed design of this program will be presented by a forthcoming paper from the collaboration.
We obtained NIRCam imaging of the Frontier-Field cluster MACSJ0416.1--2403 \citep{Lotz_17} with both the F210M and F480M filters on August 20, 2023, with total exposure time 9.3\,ksec (2.58\,h) and 3.2\,ksec (0.89\,h) in the F210M and F480M bands, respectively. 
we also obtained two rows of NIRCam WFSS observations with the F480M filter and the column direction grism (Grism C), each with four dithers, with total exposure time 6.2\,ksec (1.72\,h), and the on-source  time of 3.1\,ksec (0.86\,h) for our target, MACS0416-Y3. 
All JWST data taken with this program have been made publicly available immediately on MAST\footnote{\url{https://mast.stsci.edu/}}.

\subsection{Image Reduction}
We reduce NIRCam imaging data with version 1.11.2 of the JWST calibration pipeline.
Imaging data reduction is performed through the standard stage-1/2/3 pipeline with customized steps. 
We perform the so-called ``snowball'' masking (see \citealt{Rigby_23}) using the stage-1 pipeline.
We subtract the ``1/f'' noise using the median of each row and column in stage-2. 
We also remove low-level background (including the ``wisps'') using the median-stacked image taken with each detector with proper masking of real sources.
Imaging data are mosaicked in stage-3 with a pixel size of 0\farcs03 and \texttt{pixfrac=1}.
The astrometry of mosaicked images have been registered to Gaia-DR3 \citep{Gaia_23}.

\subsection{WFSS Reduction}
\begin{figure*}
    \centering
    \plotone{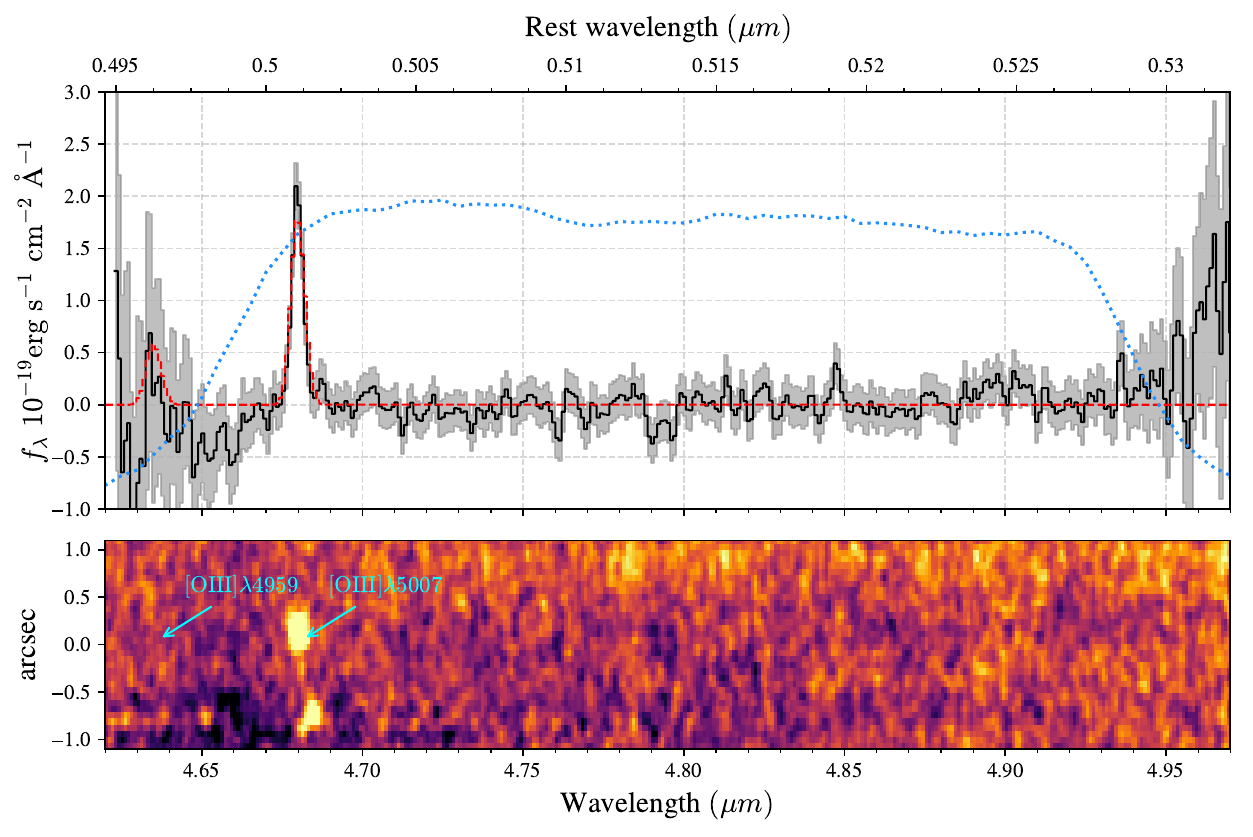}
    \caption{1D (top) and 2D (bottom) spectrum of MACS0416-Y3. The red dashed line shows the best-fit model of the 1D spectrum. We overplot the F480M sensitivity as a blue dotted line to show that the $\OIII\lambda4959$ line falls out of the F480M wavelength coverage with a non-detection, while $\OIII\lambda5007$ is securely detected with SNR=16.3. We detect the other emission line ∼ $\sim1$ arcsec below the center with $\sim190\rm km\ s^{-1}$ offset from a companion \OIII emitting galaxy of MACS0416-Y3 (the ID is MACS0416-Y2 in \citealt{Laporte_15}).}
    \label{fig:spec}
\end{figure*}
We use version 1.8.3 of the JWST Calibration pipeline \texttt{CALWEBB} Stage 1 to calibrate individual NIRCam WFSS exposures, with reference files \texttt{jwst\_1090.pmap}. The $1/f$ noise is then subtracted along rows for our Grism-C exposures using the routine described in \citet{wang_23}. The world coordinate system (WCS) information is assigned to each exposure with \texttt{assign\_wcs} step. The flat field is done with \texttt{CALWEBB} stage-2. we build the median backgrounds based on all of the MAGNIF WFSS exposures, which are then scaled and subtracted from each individual exposure. We apply an additional background subtraction, which is estimated by \texttt{photutils}\citep{larry_bradley_2022_6825092} to remove the residual background. We then measure the astrometric offsets between each of the short wavelength (SW) images and the fully calibrated F480M mosaic to align each grism exposure with the direct image. The correcting of this offset is important since the grism WCS should be aligned with direct image for the tracing model to work properly \citep[c.f.][]{Sun_2023}. 

The pre-processed WFSS exposures are then processed by Grism Redshift \& Line {Analysis tool} (\textsc{Grizli}\footnote{\url{https://github.com/gbrammer/grizli}}; \citealt{Brammer_2022}). We use the spectral tracing and grism dispersion models \citep{Sun_2023} that were produced using the JWST/NIRCam commissioning data of the Large Magellanic Cloud (LMC; PID: 1076). The sensitivity models are from JWST/NIRCam Cycle-1 absolute flux calibration observations (PID: 1536/1537/1538). Both the tracing, dispersion and sensitivity \footnote{\url{https://github.com/fengwusun/nircam_grism/}} are converted to format accepted by \textsc{Grizli}. The detection catalog for spectral extraction is built from the F480M direct image, and the continuum cross-contamination is subtracted by \textsc{Grizli} forward modeling using the F480M image as the reference image for each grism exposure. The 1D spectra are extracted with optimal extraction \citep{Horne_86}, and the emission line maps are drizzled from the 2D grism to the same WCS frame with \texttt{pixfrac=1} and pixel size=0\farcs03, corresponding to $\rm 5~\AA$ on the wavelength axis.

\section{Results and Analysis}\label{sec:analysis}
\subsection{Lens Magnification}
\label{SectionThreeTwo}

We adopt the publicly available lens model of the MACSJ0416--2403 field which was constructed by \citet{Richard_21} using the \texttt{Lenstool} \citep{Jullo_07} public software and constrained with a large number of spectroscopically confirmed multiple systems with VLT/MUSE.
At this location and source redshift $z_{\rm spec}=8.34$, the model predicts a lensing magnification factor of $\mu=1.49$, which is consistent with the magnification computed with other publicly available lens models ($\mu\simeq1.4-1.7$; e.g., \citealt{Diego_15}, \citealt{Jauzac_15}, \citealt{Zitrin_15}, \citealt{Kawamata_16}, \citealt{Okabe_20}), suggesting a limited uncertainty ($\lesssim10\%$) from the lens model. While Diego et al. (in prep) suggest higher magnification $\mu=2.2$ from the latest JWST model, we utilize $\mu=1.49$ for this paper to be more consistent with current models in literature.
We apply the deflection map from the \texttt{Lenstool} model to reconstruct the source-plane image of MACS0416-Y3 in the F480M band, together with the corresponding \texttt{WebbPSF} \citep{Perrin_14} model of the F480M PSF.
The reconstructed image is shown in the second panel of Figure \ref{fig:dir}.

\subsection{Physical Properties of MACS0416-Y3}\label{sec:prop}
The redshift of MACS0416-Y3 was previously determined as $z_{\rm phot}=9.3^{+0.4}_{-0.5}$ (\citealt{Laporte_15}; see also \citealt{Coe_15}, \citealt{McLeod_15}, \citealt{Infante_15}, \citealt{Merlin_16}), this object being an HST $Y_{105}$-band dropout. 
Combining deep HST/ACS photometry \citep{Lotz_17, Steinhardt_20} and eight-band JWST photometry at 0.8-5.0\,\micron\ from PEARLS program (\citealt{Windhorst_23}; N.\ Foo et al.\ in prep.), we derive a lower photometric redshift $z_{\rm phot}=8.95^{+0.05}_{-0.14}$. 
The bright emission line with $\sim16\sigma$ detection at 4.679\,\micron\ in the F480M grism data (Figure~\ref{fig:spec}) is $\OIII\lambda5007$ , yielding a spectroscopic redshift for MACS0416-Y3 of $z_{\rm spec}=8.343^{+0.002}_{-0.002}$.
Solutions where this line is either $\OIII\lambda4959$ or H$\beta$ can be easily ruled out because the brighter $\OIII\lambda5007$ would also enter the F480M bandwidth.


We model the spectral energy distribution (SED) of MACSJ0416-Y3 using all available JWST/NIRCam photometry (N.\ Foo et al. in prep) with \texttt{BAGPIPES} \citep[Bayesian Analysis of Galaxies for Physical Inference and Parameter EStimation;][]{Carnall_18, Carnall_19}, adopting the default \texttt{BAGPIPES} stellar population models \citep[][]{Chevallard_16}, which are the 2016 updated version of the models from \citet[][]{Bruzual_03}, assuming a \citet[][]{Kroupa_01} initial mass function (IMF). These models were updated to include the stellar spectral library from MILES \citep[][]{Falcon-Barroso_11} alongside the most recent stellar evolutionary tracks from PARSEC \citep[][]{Bressan_12} and COLIBRI \citep[][]{Marigo_13}. We use the default \texttt{BAGPIPES} nebular emission models which are constructed following the methodology of \citet[][]{Byler_17} using the 2017 updated version of the \texttt{Cloudy} photoionization code \citep[][]{Ferland_17}. The metallicity of the ionized gas is assumed to be the same as that of the stars used to produce the ionizing photons. Finally, we assume the \citet[][]{Calzetti_00} attenuation model. 
A constant star formation history (SFH) is assumed. We fix the redshift at the spectroscopic value for MACS0416-Y3. Log-uniform priors are assumed for the stellar mass in the range $5 < \mathrm{log}_{10}\left(M_{\ast}/M_{\odot}\right) < 13$ and the stellar age in the range $1\,\mathrm{Myr} < t_{\ast} < t_{\mathrm{univ}}$, where $t_{\mathrm{univ}}$ is the age of the Universe at the observed spectroscopic redshift. Uniform priors are assumed for the stellar metallicity in the range $-2.0 < \mathrm{log}_{10}\left(Z_{\ast}/Z_{\odot}\right) < +0.5$, the V-band dust attenuation in the range $0 < A_{V} < 8$, and the ionization parameter in the range $-4 < \mathrm{log}_{10}\left(U\right) < -2$.

The results of the \texttt{BAGPIPES} fitting suggest that MACS0416-Y3 is a star-forming galaxy with a stellar mass of $10^{8.6 \pm 0.1}\,M_{\odot}$ and a star-formation rate averaged over the last $10\,\mathrm{Myr}$ of $42 \pm 5\ M_{\odot}/\mathrm{yr}$. These values are all corrected by the lensing magnification factor $\mu=1.49$ as described in Section~\ref{SectionThreeTwo}. In addition, MACS0416-Y3 is found to have a young stellar population with a mass-weighted age of $5.3^{+0.5}_{-0.2}\,\mathrm{Myr}$ and moderate amounts of dust attenuation ($A_{V} = 1.0 \pm 0.1\,\mathrm{mag}$). Furthermore, we compared the \texttt{BAGPIPES} results to those from \texttt{Prospector} \citep{Johnson_21}, which provides similar functionality but differs in numerous modeling aspects. Following the methodology outlined in \citet{Tacchella_22}, the results of the \texttt{Prospector} fitting suggest that MACS0416-Y3 is sightly less massive ($10^{8.3 \pm 0.1}\,M_{\odot}$). The two measurements are consistent within estimated errors, and we adopt \texttt{BAGPIPES} model, which better reproduces the observed SED with a smaller chi-square.


\subsection{Dynamical Modeling}\label{sec:dyn}
\begin{figure*}
    \epsscale{1.15}
    \centering
    \plotone{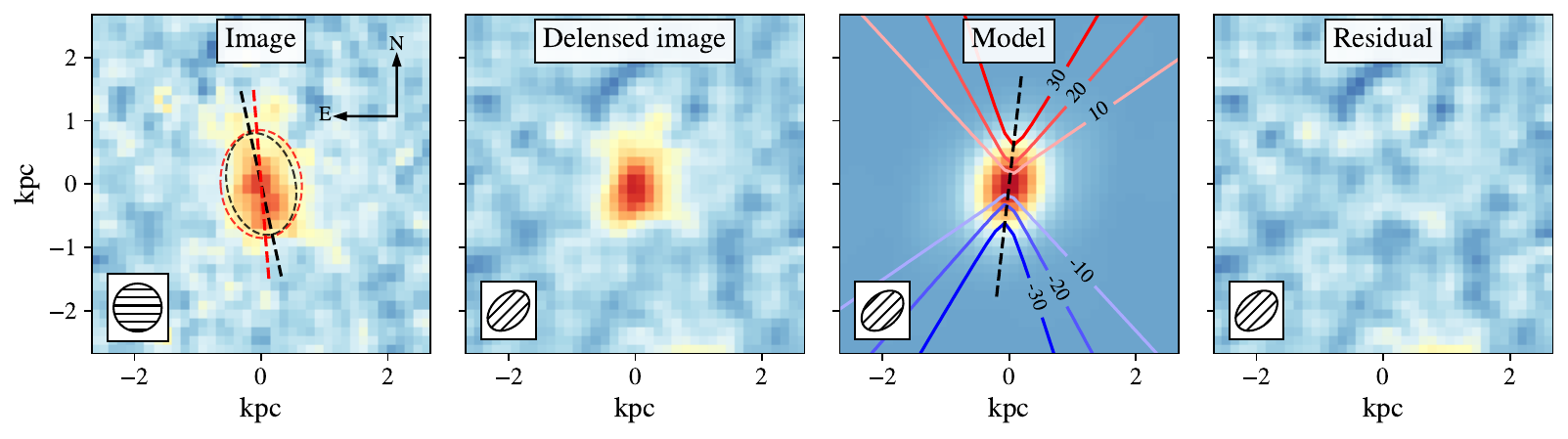}
    \caption{From left to right: the F480M direct image; reconstructed image after lensing correction; best-fit S\'{e}rsic model with PSF convolved; and the fit residuals. The original and source plane reconstructed PSFs are shown on left bottom of the corresponding panels. In the first panel, we overplot the S\'ersic models of the direct image (black dashed ellipse, with black dashed line the major axis) and \OIII map (red dashed ellipse, with red dashed line the major axis) to show their spatial offset (i.e., hint of rotation) that can be directly observed. In the third panel, the contours represent the model velocity field in units of $\rm km\ s^{-1}$, and the black dashed line is the rotation axis. The North and East orientations of all the images are indicated by the black compass in the first panel.}\label{fig:dir}
\end{figure*}

The basic method of grism dynamical modeling has been proposed by \citet{Outini_20} and applied on HST WFC3 grism spectra. Recently, \citet{Graaff_23} applied dynamical modeling on JWST NIRSpec MSA spectra. Here we present the technique on JWST NIRCam grism.

To obtain the dynamical modeling of the $\OIII\lambda5007$ emission of MACS0416-Y3, the procedures are summarized as follows: 

\begin{enumerate}
    \item We first fit the 2D surface brightness profile to the F480M direct image in the delensed source plane.
    \item Using the derived morphological parameters, we generate the velocity field in the source plane.
    \item We then map the source plane velocity information to the image plane with the lens model from \citet{Richard_21}.
    \item We use the F480M direct image as reference image and convolve it with the rotation velocity and velocity dispersion field from step 2 in the image plane.
    \item Finally, we run Markov chain Monte Carlo (MCMC) sampling to derive the best-fit kinematic parameters and their uncertainties.
\end{enumerate}

We fit a S\'{e}rsic profile convolved with the lensing reconstructed PSF to the F480M source-plane direct image using \texttt{PetroFit} \citep{Geda_2022}. 

Assuming an intrinsic circular disk, the inclination angle can be expressed as $\cos^2(i)=[(b/a)^2-q_0^2]/(1-q_0^2)$ (i.e., $i=90\degree$ for an edge-on galaxy, see \citealt{Hubble_1926}), where $b/a$ is the axis ratio of the 2D surface brightness profile, and $q_0$ is the intrinsic axis ratio defined by the third axis over long axis. For simplicity, we assume $q_0=0$. 
We note that the projected axis ratio can also be produced by thick disks which have been found common in high redshift $z\gtrsim1-2$ \citep{Zhang_19} even up to $z\sim8$, 
i.e., 
with an observed ratio $b/a\sim0.6$ in our case, a thick disk with $q_0=0.5$ would increase $\sin(i)$ by $\sim15\%$  and influence the velocity in Eq (\ref{eq:2}). Thus, the assumption of a thin disk could lead to an rotation velocity uncertainty up to $\sim15\%$.

We also fit the position angle of the galaxy, which is later used to determine the position of the velocity field. Figure \ref{fig:dir} shows the best-fit results and three parameters of position angle (PA), effective radius ($R_e$) and inclination angle ($i$) are listed in Table \ref{tab:info}.

We use an arctangent profile model \citep{Neeleman_20} for rotation curve parameterized as:
\begin{equation}\label{eq:1}
    v(R)=\frac{2}{\pi}v_{\rm rot}\arctan(\frac{R}{R_v})+v_0,
\end{equation}
where $R$ is the galactocentric distance, $v_{\rm rot}$ is the maximum rotation velocity, $R_v$ is a scaling factor determining the steepness of the rotation curve, and $v_0$ is the constant systemic velocity. The systemic velocity should be zero if the redshift is secured, but we keep this parameter free in our model to compensate for the small redshift uncertainty in the fitting. The line-of-sight velocity can be expressed as:
\begin{equation}\label{eq:2}
    V_{\rm los}(x,y)=v(R)\cos(\theta)\sin(i),
\end{equation}
where $x,y$ are the Cartesian coordinates in the sky plane, $\theta$ is the polar angle in the galaxy source plane, and $i$ is the inclination angle of the galaxy. To relate the galaxy plane to the sky plane, we have that: $\tan(\theta)=\tan(\phi)/\cos(i)$,
where $\phi$ is the polar angle in the sky plane. 

The gravitational lensing distorts trajectories of photons but leaves frequencies unchanged, and the observed velocity field is also sheared \citep{Xu_23}. We convert the source plane velocity given by Eq (\ref{eq:2}) to the image plane using the deflection map obtained with the lens model, similar to that in \citet{Tokuoka_22}.

We use the F480M image as the reference to model the \OIII line in the 2D grism. Since the F480M image also includes photons from the continuum and the SED modeling estimates that $\sim57\%$ of the flux density in F480M is contributed by the $\OIII\lambda5007$, we thus add a normalization parameter $s$, defined as the ratio between the $\OIII\lambda5007$ flux and the total flux in F480M band, to rescale the model flux. In addition, this scaling factor automatically takes the uncertainty in photometric zero points into account. An alternative way is to use a neighboring filter to subtract continuum flux from F480M, however, considering extra noise this may introduce, we apply the former.

The original F480M mosaicked image and corresponding grism $\OIII\lambda5007$ map are aligned such that North is up and East is to the left, and we rotate the cutout by the position angle of the telescope so that the spectral dispersion direction increases towards the right. In order to conveniently  model the dispersed 2D emission lines. 
we first convolve a 1D Gaussian kernel with a variable sigma that corresponds to the spectral resolution ($R\sim1650$ at 4.68\,\micron\footnote{\url{https://jwst-docs.stsci.edu/jwst-near-infrared-camera/nircam-instrumentation/nircam-grisms}}) along the wavelength direction for each row.
The grism image is smoothed in a similar manner to suppress noise fluctuation below the resolution.  After convolution, the image PSF is matched to the grism line spread function (LSF). Given an observed rotation velocity field, the reference image can be dispersed into the $\OIII\lambda5007$ emission line map on the grism frame. Assuming the direct image is dominated by \OIII flux or that the \OIII is relatively smooth and follow the stellar distribution, the model emission line map can be expressed as the convolution of the direct image with the rotation velocity field. With zero velocity, the direct image can be directly transformed to the grism emission map by just changing the $x$ coordinate unit from arcsec to wavelength (for a pixel scale=0\farcs03 in the spatial direction, the corresponding pixel scale in wavelength is $5\rm \AA$). For a non-zero velocity field $V_{\rm los}(x,y)$, every pixel in the direct image at coordinate ($x_0,y_0$) has a wavelength offset applied:
\begin{equation}
    \Delta\lambda(x_0,y_0)=\frac{V_{\rm los}(x_0,y_0)}{c}\lambda_{\rm obs},
\end{equation}
where $\lambda_{\rm obs}$ is the observed wavelength of the $\OIII\lambda5007$ emission line in our case, and $c$ is the speed of light. The location of each dispersed pixel is then expressed as: $(x',y')=(x_0+\Delta\lambda/(5\rm \AA),y_0)$. We note that here we neglect the tilt of the spectral tracing, as the spatial offset in pixel $dy$ is $\sim0.001$ of the offset in wavelength direction $dx$ \citep{Sun_2023}. 
Given the size of the $\OIII\lambda5007$-emitting region ($dx\sim10$), the spatial offset $dy$ is negligible. 

\begin{figure*}
\epsscale{1.15}
    \centering
    \plotone{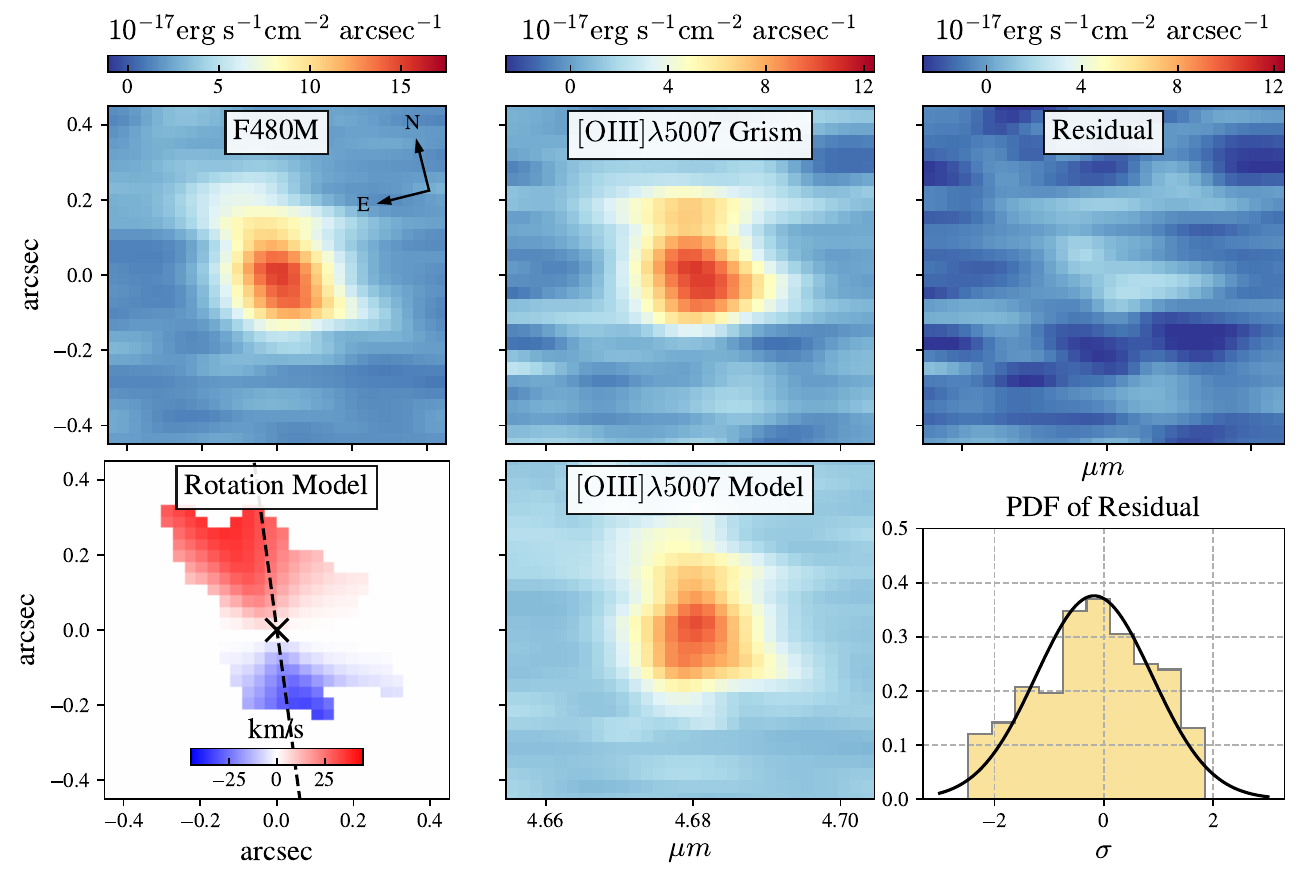}
    \caption{The first column shows the direct F480M image (top), the best-fitted rotation velocity field (bottom), with the black dashed line representing the major axis fitted in the source plane and the black $\times$ marking the model kinematic center. Second column shows the $\OIII\lambda5007$ line map (top), and the $\OIII\lambda5007$ model constructed from reference image (bottom). Both the reference image and line map are on the image plane and aligned such that the spectrum dispersion is left to right, and the North and East are annotated by the black compass. The last column shows the residual of subtracting model from $\OIII\lambda5007$ line map relative to background noise $\sigma$ (top), and histogram of residuals (bottom). The residuals are well fitted by a Gaussian $\mathcal{N}(-0.18,1.06)$, as shown in black curve.}
    \label{fig:model}
\end{figure*}

The shifted model is then resampled to the same pixel grid using the Cloud-in-Cell (CIC) algorithm \citep{Birdsall_69} to conserve photons. To model the broadening due to the velocity dispersion, we convolve a 1-dimensional Gaussian kernel along each row with: 
\begin{equation}
    \sigma=\frac{v_\sigma}{c}\lambda_{\rm obs},
\end{equation}
where $v_\sigma$ is the velocity dispersion. For simplicity, we assume that $v_\sigma$ is a constant over the disk.

The five free parameters in our model are rotation velocity $v_\mathrm{rot}$,  rotation curve scaling factor $R_v$, systemic velocity $v_0$, velocity dispersion $v_\sigma$ and the model scaling factor $s$. We keep the other parameters fixed from the S\'{e}rsic fitting previously described. We explore the model parameter space using the MCMC sampler $\textsc{Emcee}$ package \citep{2013PASP..125..306F}. The likelihood function is defined as $L\propto \exp(-\chi^2/2)$ with:
\begin{equation}
    \chi^2 = \sum_{x,y}\frac{(M_{\rm obs}(x,y)-M_{\rm model}(x,y))^2}{\sigma_{\rm obs}^2(x,y)},
\end{equation}
where $M_{\rm obs}, M_{\rm model}$ are the observed line map and model, $\sigma_{\rm obs}$ is the uncertainty of measured surface brightness. 

The prior and posterior of the MCMC sampling are discussed in Appendix \ref{sec:mc}. The best-fit parameters are listed in Table \ref{tab:info}. Figure \ref{fig:model} shows a comparison between the observed and best-fit \OIII line maps in the image plane. The residuals relative to background noise follow a Gaussian distribution $\mathcal{N}(-0.18,1.06)$, with mean -0.18 and standard deviation 1.06, indicating a good fit.

One caveat is that the inferred velocity dispersion $v_\sigma=40^{+30}_{-26}\ \rm km\ s^{-1}$ falls below the grism velocity resolution $\sigma\approx76\ \rm km\ s^{-1}$, which provides an upper limit for this parameter, and we caution the use of this value until future follow-up confirmation. Even so, our measured velocity dispersion is comparable to the typical velocity dispersion at $z\gtrsim6$ measured with ionized gas \OIII and \Ha \citep{Graaff_23}.

To test whether rotation is needed to reproduce the observed grism spectrum, we compare our best-fit model with a no-rotation model, the latter having zero rotation velocity while keeping other parameters the same. We find that the reduced chi-square decreases by $\sim0.1$ compared to the no-rotation model, as listed in Table \ref{tab:info}. Moreover, we applied the chi-square goodness of fit test and in the null hypothesis that the observations can be described by the model, the p-values for the chi-squares of the best-fit model and the no-rotation model are 0.02 and 0.002, respectively. Assuming a significance level of 0.01, the null hypothesis can be rejected for the no-rotation model. The solution with rotation is apparently a better fit.

Nevertheless, although our model 
is an excellent fit to the current data, alternative scenarios like mergers or outflows cannot be completely ruled out. 
A merger with two or more \OIII clumps could exhibit velocity field similar to rotation, which is challenging to be distinguished \citep{Simons_19}. Some works  tried to classify major mergers and rotation disks \citep{Shapiro_08,Rizzo_22}, which usually require quantities measured at individual spaxel level in IFU datacubes to quantify certain asymmetric parameters. However, it is unpractical to extract detailed information for individual pixels in overlapped 2D spectra, and we are unable to model such complex asymmetries in our assumption of a simple rotation disk. In addition, there is a companion galaxy MACS0416-Y2 to the South of MACS0416-Y3 (see Figure \ref{fig:spec}), so we cannot rule out this as a merging system.
On the other hand, the bipolar outflow or inflow could also show velocity components \citep[e.g.,][]{Barreto_19}, while we find no obvious broad components as a sign of outflows in 1D/2D spectra in Figure \ref{fig:spec} and there is no current evidence indicating a powerful active galactic nucleus that can power a bipolar outflow throughout the entire galaxy. 
Nevertheless, considering the measured rotation velocity with relatively high uncertainty $v_{\rm rot}=58^{+53}_{-35}\rm\ km\ s^{-1}$, we report this system as a tentative rotating disk at the present stage. The following analysis is therefore based on the rotation assumption. 

\begin{deluxetable}{cc}
\tablecaption{Physical properties of MACS0416--Y3.}
\tablehead{\colhead{Property} & \colhead{Value}}
\startdata
\multicolumn{2}{c}{Basic Properties}\\
R.A. (deg) & 64.048125 \\
Dec. (deg) & -24.081452 \\
$z_{\rm spec}$ & $8.343\pm0.002$\\
$f_{\OIII\lambda5007}$ ($\rm erg\ s^{-1} cm^{-2}$) & $(1.03\pm0.06)\times10^{-17}$\\
$\mu$ & 1.49 \\
\hline
\multicolumn{2}{c}{Morphology Properties\tablenotemark{a}}\\
$R_e$ (kpc) & $0.29\pm0.01$\\
PA (deg) & $83.6\pm5.1$\\
$i$ (deg) & $53.1\pm2.3$\\
\hline
\multicolumn{2}{c}{SED Properties \tablenotemark{b}}\\
$\log(M_*/M_\odot)$ & $8.6^{+0.1}_{-0.1}$\\
SFR ($M_\odot\ \rm yr^{-1}$)& $42^{+5}_{-5}$ \\
\hline
\multicolumn{2}{c}{Best-fit Model Parameters} \\
$v_{\rm rot}\ (\rm km\ s^{-1})$& $58^{+53}_{-35}$ \\
$v_\sigma\ (\rm km\ s^{-1})$& $40^{+30}_{-26}$\\
$v_0\ (\rm km\ s^{-1})$& $-3^{+12}_{-12}$ \\
$R_v$ (kpc) & $0.62^{+0.28}_{-0.27}$ \\
$s$ & $0.63^{+0.04}_{-0.04}$ \\
reduced $\chi^2_{\rm Best}$\tablenotemark{c} & 1.18 \\
reduced $\chi^2_{\rm no-rot}$\tablenotemark{c} & 1.27 \\
\hline
\multicolumn{2}{c}{Derived Properties} \\
$v_{\rm rot}/v_\sigma$ & $1.6^{+1.9}_{-0.9}$\\
$\log(M_{\rm dyn}/M_{\odot})$ & $8.4^{+0.5}_{-0.7}$\\
\enddata
\tablenotetext{a}{Values fitted on source plane by PetroFit. The orientation of PA is defined such that right is 0\degree and up is 90\degree.}
\tablenotetext{b}{These values are corrected by lens magnification $\mu=1.49$.}
\tablenotetext{c}{$\chi^2_{\rm Best}$ is calculated with our best-fit model, while $\chi^2_{\rm no-rot}$ assumes no rotation for comparison.}

\end{deluxetable}\label{tab:info}
\section{Discussion} \label{sec:res}


\subsection{A possible quasi-stable rotating disk at $z\gtrsim8$}\label{sec:stable_disk}

\begin{figure}[t]
\epsscale{1.2}
    \centering
    \plotone{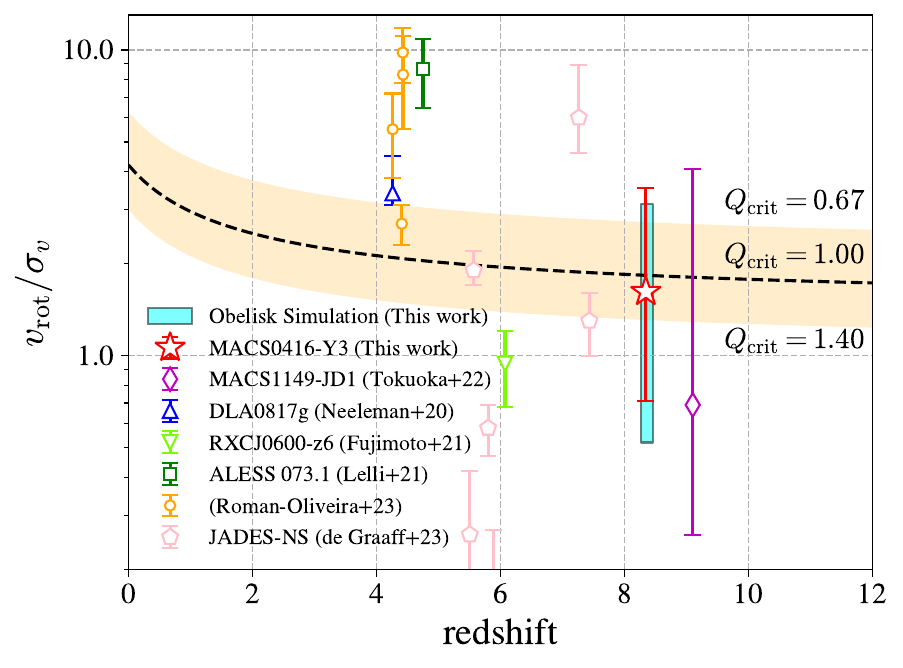}
    \caption{Kinematic ratio $v_{\rm rot}/\sigma_v$ as a function of redshift. The red star shows our result with NIRCam WFSS, and other symbols show literature observations using ALMA \citep{Neeleman_20,Lelli_21,Tokuoka_22,Fujimoto_21,Oliveira_23} and JWST NIRSpec MSA \citep{Graaff_23}. The cyan shadowed area represents the $1\sigma$ interval predicted by \textsc{Obelisk} simulation. The dashed line shows the semi-empirical models based on Toomre’s disk instability parameter with $Q_{\rm crit}$ = 1.0, and the shadowed area represents the interval between $Q_{\rm crit}$ = 0.67 and $Q_{\rm crit}$ = 1.4 \citep{Wisnioski_15}. We extrapolated the model down to a stellar mass $\log(M_*/M_\odot)=8.6$ to compare with MACS0416-Y3.}
    \label{fig:kratio}
\end{figure}

In Figure \ref{fig:kratio}, we compare the kinematic ratio $v_{\rm rot}/\sigma_v=1.6^{+1.9}_{-0.9}$ from our dynamical modeling to literature observations \citep{Neeleman_20,Lelli_21,Tokuoka_22,Oliveira_23,Fujimoto_21,Graaff_23}. We also compare the \textsc{Obelisk} simulation \citep{Trebitsch_21} to our observation. The simulated galaxies are at $z=8.36$ with 16th and 84th percentile of stellar mass being $10^{8.4}M_\odot$ and $10^{9.3}M_\odot$. The rotation velocity is measured as the average over the galaxy of the tangential component of gas velocity, and the velocity dispersion is measured as the quadratic sum of the velocity components. Both averages are weighted by the gas density squared, as the [OIII] line emissivity. The measured kinematic ratio ranges from $\sim0.5$ to $\sim3$. Our observed kinematics ratio reasonably agrees with the simulation.

The redshift evolution of the kinematic ratio can be described by a semi-empirical model based on the Toomre disk instability parameter $Q_{\rm crit}$ \citep{Wisnioski_15}. This parameter, mainly parameterized with gas surface density and epicyclic frequency \citep{Toomre_64}, determines if a differentially rotating system is stable. 
Disks are believed to be stable against collapse above 
$Q_{\rm{crit}}$. 
We extrapolate this semi-empirical model to $z>8$ and $\log(M_*/M_\odot)=8$, 
originally fitted from a sample with $0.7<z<2.7$ and $9.2<\log(M_*/M_\odot)<11.2$. With hydrodynamical simulations, \citet{Kim_07} found $Q_{\rm crit}=0.67$ for  a thick disk with gas and stars, and $Q_{\rm crit}=1.27$ for a thin disk with gas and stars, while a thin gaseous quasi-stable disk has $Q_{\rm crit}=1$. In Figure~\ref{fig:kratio}, we show this model prediction within $0.67<Q_{\rm crit}<1.4$, and find that our results appear consistent with semi-empirical model, in preference of a quasi-stable gaseous disk. 

The gas fraction (atomic gas and molecular gas) of the disk can be estimated as: $f_{\rm gas} = M_\mathrm{gas} / (M_\mathrm{gas} + M_\mathrm{star})= (a/Q_{\rm crit})(\sigma_v/v_{rot})$, where $a$ depends on the 
velocity profile, e.g., $a=1$ for a Keplerian disk and $a=2$ for a solid-body disk \citep{Genzel_11}. Taking $Q_{\rm crit}=1$, the modelled kinematic ratio then indicates a gas fraction $f_{\rm gas}\gtrsim0.6$. This is consistent with high redshift main sequence galaxies with high gas fraction (\citealt{Aravena_20,Walter_22}; Li et al. in prep). 
The gas fraction together with the rotation-dominated nature may indicate ongoing gas accretion, and the rotation is sustained by angular momentum from inspiraling gas \citep{Stewart_11,Stewart_17, Zhang_23}.

\subsection{Dynamical mass versus stellar mass}

Assuming a spherical mass distribution, the dynamical mass at radius $R$ can be estimated as: 
\begin{equation}
    M_{\rm dyn}/M_\odot=2.32\times 10^5v(R)^2R,
\end{equation}
where $v(R)$ is the rotation velocity ($\rm km\ s^{-1}$) at radius $R$ (kpc). We assume the extent of the galaxy to be three times the effective radius ($R_e=0.29$ kpc) in F480M to estimate the dynamical mass, corresponding to the region enclosing $80\%-90\%$ of the total flux density in our s\'ersic model, which is a reasonable choice to define galaxy size \citep{Neeleman_20}. We measure a dynamical mass $\log(M_{\rm dyn}/M_{\odot})=8.4^{+0.5}_{-0.7}$. As noted by \citet{Neeleman_20}, this underestimates the dynamical mass by up to $30\%$ for an exponential thin-disk instead of spherical mass distribution, due to their different rotation curves \citep{Walter_97}. Moreover, since the rotation velocity can be reduced in the presence of pressure gradients of turbulent gas \citep{Burkert_10}, the actual dynamical mass can be several times larger (e.g., \citealt{Tokuoka_22} estimated it to be $\sim5$ times larger in case of $v_{\rm rot}/v_\sigma\sim1$). With our measured kinematic ratio $v_{\rm rot}/v_\sigma\sim1.6$, we estimate the dynamical mass to be increased by a factor of $\sim 3$ (see \citealt{Tokuoka_22,Burkert_10} about this correction). The corrected dynamical mass is $\log(M_{\rm dyn}/M_{\odot})\sim8.9$. Comparing to our measured stellar mass $\log(M_*/M_\odot)=8.6^{+0.1}_{-0.1}$ (lensing corrected) and the estimated gas fraction $\gtrsim0.6$, the corrected dynamical mass broadly agrees with the sum of stellar and gas components. As this source has a young mass-weighted age $\sim5$~Myr (discussed in Section \ref{sec:prop}), the dynamical mass can be attributed to a young stellar population and high fraction of gas, and the uncertainties in stellar mass and dynamical mass also give room for dark matter to exist in this galaxy.

\subsection{Rotating disks as a consequence of gas accretion?}
The star-formation history of high redshift galaxies could be dominated by continuous or bursty star-formation, with a much higher gas fraction than today \citep{Tacconi_20}. Assuming the gas fraction $f_{\rm gas}=0.6$ for MACS0416-Y3 as inferred in Section \ref{sec:stable_disk}, 
the gas mass is estimated to be $M_{\rm gas}\sim10^{8.8}\,M_\odot$. With the star formation rate $42\,M_\odot/\rm yr$ from SED fitting, this leads to a short gas depletion time $t_{\rm dep}\sim15\,\rm Myr$. 
This high sSFR should be sustained by continuous gas replenishment via cold mode accretion \citep{Dekel_09,keres_09,Cresci_10}, otherwise it would be quenched at $z\sim8.19$. The cold streams tend 
to orbit with high angular momentum before building the 
galactic disk \citep{Stewart_11, Zhang_23}. 
In this scenario, we would 
expect the gas and young stars to settle into an ordered rotation. 
Therefore, the gas with high angular momentum in our observed disk at $z\gtrsim8$ is possibly brought from the efficient accretion at early stage of galaxy mass assembly \citep{Neeleman_20,Tacchella_23,Heintz_23}. Further studies are still needed to investigate the coupling between accreted gas and galaxy disks.

\section{Conclusions} \label{sec:con}
Through forward modeling of the JWST NIRCam grism spectra of MACS0416-Y3, we identify a possibly rotation-dominated disk at $z=8.34$. The kinematic ratio of this galaxy is within the expectation from a semi-empirical model. We conclude that this source may have already built a quasi-stable gas disk with stellar component and ongoing star formation from cold gas. 
Our discovery and similar ones from the literature \citep{Tokuoka_22} suggest that it is possible for disks to form at $z\gtrsim8$, as predicted in some simulations like Aspen \citep{Katz_19} and Obelisk. These findings make it viable to study the formation history of galaxies starting from the first billion years of the Universe \citep{Xiang_22}.

Moreover, our work illustrates the possibility for JWST NIRCam WFSS to study galaxy kinematics even at $z>8$. The forward modeling technique proposed in this letter can also be applied to galaxies with other bright optical or near infrared lines (e.g., \Ha, $\rm P\alpha$, $[\textrm{S}~\textsc{iii}]$). Given the large field of view of the NIRCam grism (up to $\sim9$\,arcmin$^2$, in contrast to 9\,arcsec$^2$ of that for NIRSpec integral field unit, IFU spectroscopy), we expect that a larger sample of disks can be discovered soon in other current and future WFSS surveys, e.g., ASPIRE \citep{wang_23}, FRESCO \citep{Oesch_23}, EIGER \citep{Kashino_23} and so on. 
As an extra example, we also present the dynamic forward modeling of a luminous \Ha-emitting galaxy at $z=5.39$ in FRESCO (\citealt{Nelson_23}) in Appendix \ref{sec:extra}, where the rotation disk can be easily modeled. The NIRCam WFSS can be a powerful tool to pre-select high-redshift galaxies with observable kinematics for future detailed follow-up studies. 

This is a pathfinder study to forward modeling the JWST NIRCam WFSS emission-line spectra, and therefore the possible systematic errors are still under investigation. We note the complexity of slitless spectroscopy where the self-contamination and cross-contamination have been a long-term problem, and the spectra overlapping may impact any measurement.  Results presented in this letter can be validated by future observations with JWST NIRSpec IFU observations and high-resolution ALMA imaging, with kinematics from both ionized gas with $\OIII\lambda5007$ and relatively cooler gas with \cii.

This grism forward modeling code is under 
development and will be made publicly available in our future 
work based on a larger sample. 

\section{Acknowledgement}
This work is based on observations made with the NASA/ESA/CSA James Webb Space Telescope.  The data were obtained from the Mikulski Archive for Space Telescopes at the Space Telescope Science Institute, which is operated by the Association of Universities for Research in Astronomy, Inc., under NASA contract NAS 5-03127 for JWST. These observations are associated with program \#2883, 1895, 1176.
The authors sincerely thank the FRESCO team (PI: Pascal Oesch) for developing their observing program with a zero-exclusive-access period.
This work made use of the High Performance Computing resources at Tsinghua University.

ZL, ZC, XL, ML, YW and SZ are supported by the National Key R\&D Program of China (grant no. 2018YFA0404503), the National Science Foundation of China (grant no. 12073014), the science research grants from the China Manned Space Project with No. CMS-CSST2021-A05, and Tsinghua University Initiative Scientific Research Program (No. 20223080023). ZL thanks Shiwu Zhang for discussions and emotional support through the work.
FS, JMH,  EE, CC, CNAW acknowledges JWST/NIRCam contract to the University of Arizona NAS5-02015.
MT acknowledges support from the NWO grant 0.16.VIDI.189.162 (``ODIN''). 
MO acknowledges the support by JSPS KAKENHI Grant Numbers JP22H01260 and JP22K21349.
KK acknowledges the support by JSPS KAKENHI Grant Numbers JP17H06130 and JP22H04939.
RAW acknowledges support from NASA JWST Interdisciplinary Scientist grants
NAG5-12460, NNX14AN10G and 80NSSC18K0200 from GSFC.
GEM acknowledges the Villum Fonden research grant 13160 ``Gas to stars, stars to dust: tracing star formation across cosmic time'' grant 37440, ``The Hidden Cosmos'', and the Cosmic Dawn Center of Excellence funded by the Danish National Research Foundation under the grant No. 140. 
AZ acknowledges support by Grant No. 2020750 from the United States-Israel Binational Science Foundation (BSF) and Grant No. 2109066 from the United States National Science Foundation (NSF); by the Ministry of Science \& Technology, Israel; and by the Israel Science Foundation Grant No. 864/23.

\facilities{JWST (NIRCam)}
\software{Python, astropy \citep{2018AJ....156..123A}, \texttt{CALWEBB} \citep{bushouse_howard_2023_8247246}, \textsc{Grizli} \citep{Brammer_2022}, \textsc{Bagpipes} \citep{Carnall_18}, PetroFit \citep{Geda_2022}, \texttt{Numpy} \citep{Harris_20}, \texttt{Scipy} \citep{Virtanen_20}, \textsc{emcee} \citep{2013PASP..125..306F}, Matplotlib \citep{Hunter_07}, \texttt{corner} \citep{corner}, \texttt{photutils}\citep{larry_bradley_2022_6825092}, \texttt{WebbPSF} \citep{Perrin_14}}

\appendix
\section{MCMC sampling}\label{sec:mc}
\begin{figure*}
    \centering
    \plotone{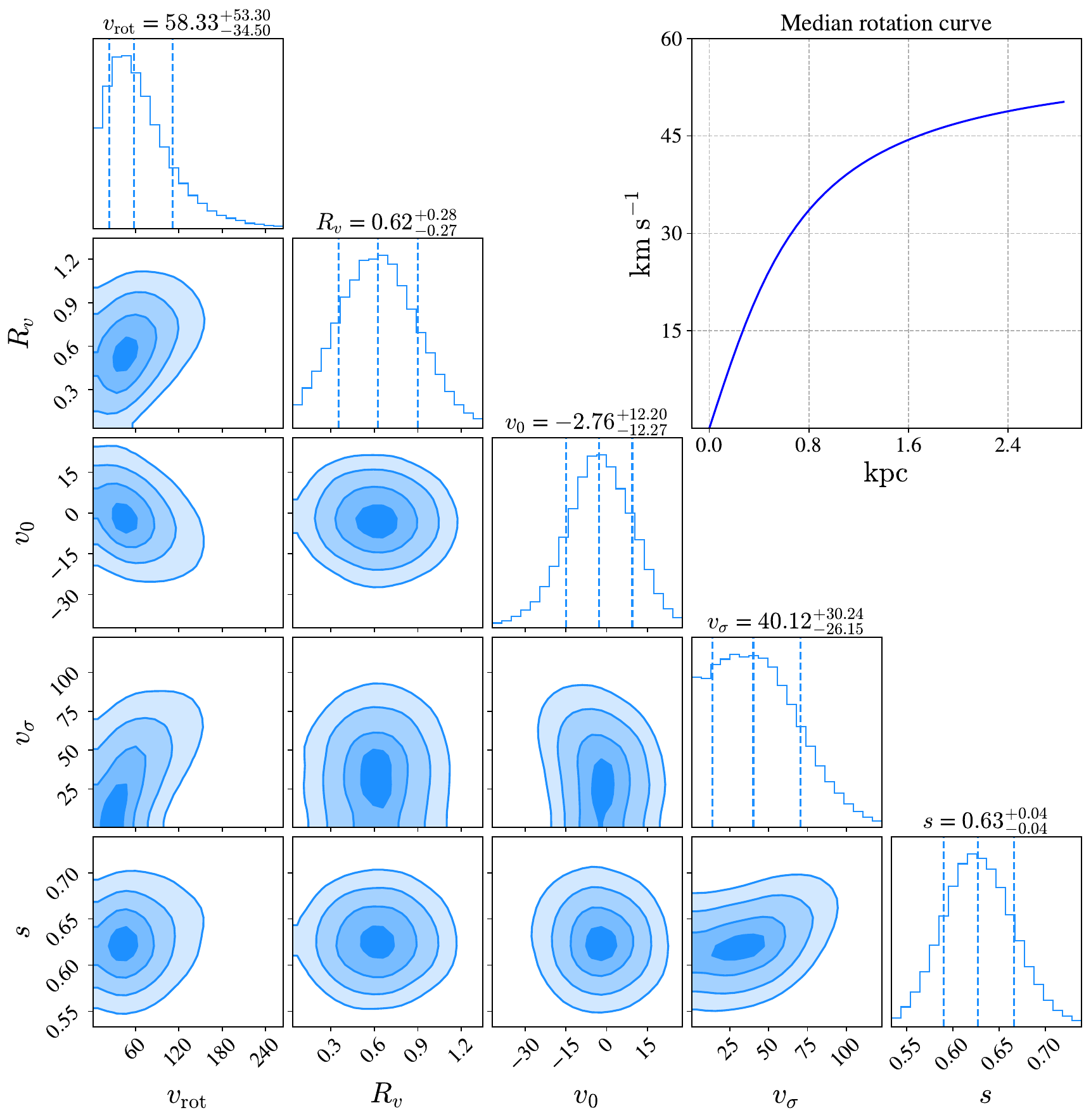}
    \caption{Posterior distribution of $v_{rot},R_v$, $v_0$ and $s$ in our dynamical model from MCMC sampling. The values on the top of each column are the medians with $1\sigma$ uncertainties. The top right panel shows the rotation curve with the median of each parameters.}
    \label{fig:mcmc}
\end{figure*}

 We apply a flat prior for the following parameters: $v_{\rm rot}\sim\mathcal{U}(0,300)\ \mathrm{km\ s^{-1}}, v_0\sim\mathcal{U}(-30,30)\ \mathrm{kms^{-1}}$, $v_\sigma\sim\mathcal{U}(0,1000) \mathrm{km\ s^{-1}}$ and $s\sim\mathcal{U}(0,100)$, and we apply a Gaussian prior on $R_v$ with $\mathcal{N}(0.5,0.3)$ kpc. The \textsc{Emcee} sampling is performed with 64 walkers, 10000 iterations each and with a burn-in period $n=200$. The posterior probability distribution for the parameters are shown in Figure \ref{fig:mcmc}, with the best-fit model shown on the top right panel. 

\section{Extra example: A remarkable rotating disk in GOODS-S field}\label{sec:extra}
We provide an extra example of a luminous $\rm H\alpha$ emitting galaxy with significant rotation signatures at $z=5.39$ (Ra=53.10171, Dec=-27.83617) found in FRESCO GOODS-S field (\citealt{Nelson_23}, \citealt{Oesch_23}; also see \citealt{Helton_23}). The FRESCO data reduction procedures are similar to Section \ref{sec:data}. We apply the same modeling procedures in Section \ref{sec:dyn}, except for lensing correction steps, because GOODS-S is not a lensing field. As this source has a stronger continuum, which has a strong impact on the flux distribution of the $\rm H\alpha$ model, we thus make the reference image from F444W subtracted by F210M to mitigate continuum flux. The best-fit results are shown in Figure \ref{fig:Ha}. We fit the rotation velocity $v_{\rm rot}=302^{+20}_{-20}\ \rm km\ s^{-1}$ and velocity dispersion $\sigma_v=118^{+7}_{-7}\ \rm km\ s^{-1}$. This rotation velocity $v_{\rm rot}$ is the maximum velocity defined in Eq (\ref{eq:1}), and if we measure the velocity at the effective radius $R_e=2.25$ kpc for comparison with \citet{Nelson_23}, we yield $v(R_e)=211^{+21}_{-18}\rm\ km~s^{-1}$. Our measured rotation velocity and velocity dispersion are fairly consistent with \citet{Nelson_23} (see their Figure 3 and Figure 4).
\begin{figure*}
\epsscale{1.15}
    \centering
    \plotone{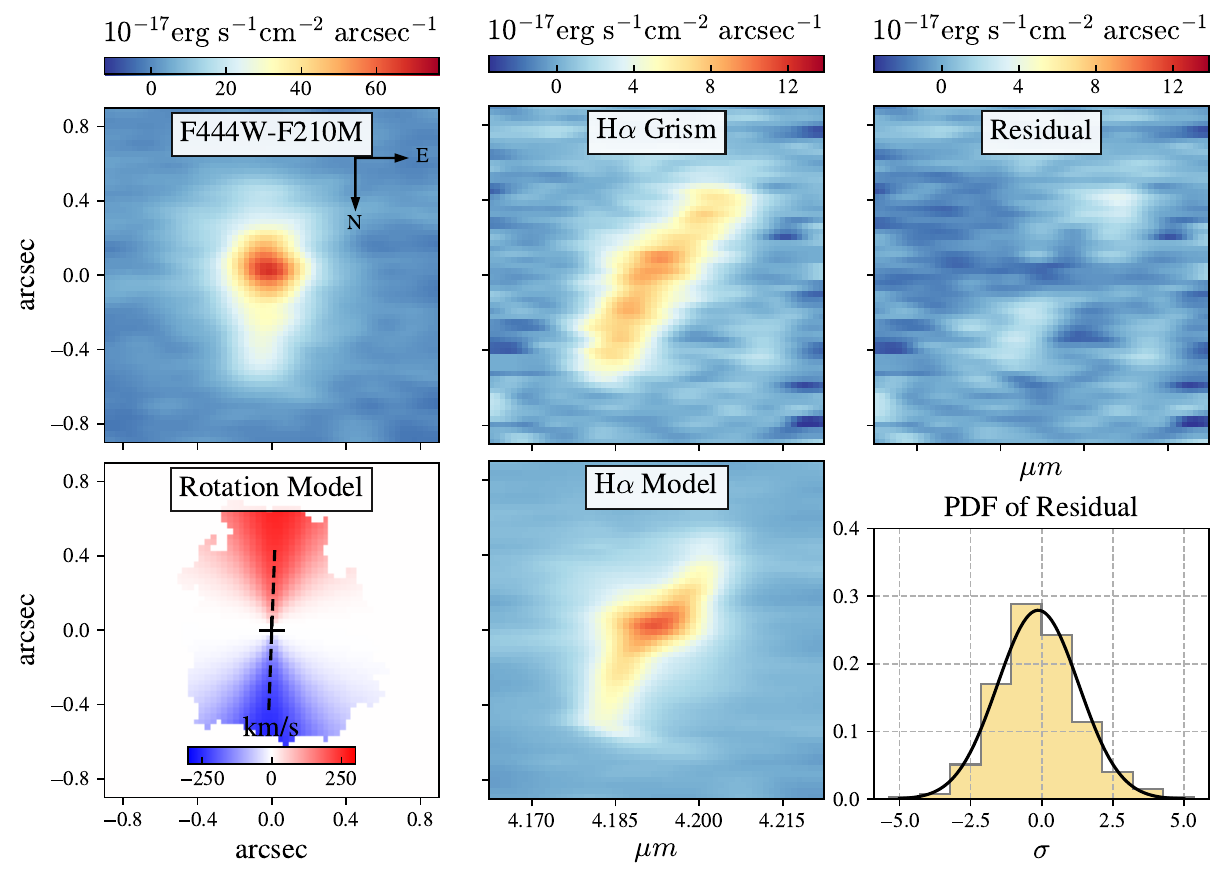}
    \caption{The same as Figure \ref{fig:model}, but for an \Ha emitting galaxy at $z=5.39$ in FRESCO GOODS-S.}
\end{figure*}\label{fig:Ha}
\bibliography{sample631}{}
\bibliographystyle{aasjournal}

\end{CJK*}
\end{document}